\documentclass [dvips, 12pt] {article} 
\usepackage {graphicx}
\usepackage {pifont}
\newcommand {\EM} {\ensuremath {\eta (548)} }
\newcommand {\OM} {\ensuremath {\omega (782)} }
\newcommand {\AM} {\ensuremath {a (980)} }
\newcommand {\FM} {\ensuremath {f (980)} }
\newcommand {\RM} {\ensuremath {\rho (770)} }

\newcommand {\DR} {\ensuremath {\Delta (1232)} }

\newcommand {\pion} {\ensuremath {{\pi^0}} }
\newcommand {\kaon} {\ensuremath {{K^0}} }

\newcommand*{\meson}[1]{\ensuremath {#1^\pm}}
 
\newcommand*{\quark}[1]{\ensuremath {#1}} 
\newcommand*{\Quark}[1]{\ensuremath {#1}-quark}

\newcommand*{\Approx}[1]{\ensuremath {\approx #1}} 
\newcommand*{\ApproxP}[1]{\ensuremath {\approx #1 \%}}

\newcommand {\cu} {\textcircled {\em \footnotesize u}}
\newcommand {\cs} {\textcircled {\em \footnotesize s}}
\newcommand {\csix} {\textcircled {\footnotesize 6}}
\newcommand {\ctwo} {\textcircled {\footnotesize 2}}
\newcommand {\csixu} {\textcircled {\footnotesize 6}$_u$ }
\newcommand {\ctwou} {\textcircled {\footnotesize 2}$_u$ }
\newcommand {\csixs} {\textcircled {\footnotesize 6}$_s$ }
\newcommand {\ctwos} {\textcircled {\footnotesize 2}$_s$ }

\newcommand {\cthreeu} {\textcircled {\footnotesize 3}$_u$}
\newcommand {\cthrees} {\textcircled {\footnotesize 3}$_s$}

\begin {document}

\begin {center}
{\Large The harmonic quarks and hadrons up to 1000 MeV}\par\bigskip 
{\large Oleg~A.~Teplov}\par\smallskip 
Institute of metallurgy and materiology of the Russian Academy of Science, 
Moscow.
\par e-mail: teplov@ultra.imet.ac.ru 
\end {center}

\begin{abstract}
The calculations and analysis of a variance have shown, 
that masses of mesons up to 1000 MeV
strongly correlate with a model spectrum of potential wells at mass
of harmonic up-quark 105.6 MeV. It is shown, that probability
of casual correspondence of the meson and model spectrums is less than
1 ppm. The harmonic quark model is applied to L=0 hadrons for the analysis 
of them structures and detection of them filled quark shells. 
The possible quark structures of all mesons up to 1000 MeV are given.
The completely neutral simple boson configurations are found 
for a threshold a proton-antiproton and the total mass of neutral pion and kaon. 
Mass relations are fulfilled with precision near 0.01\%. 
It is supposed, that new relations is connected with Higgs mechanism of mass 
formation. The final precision of calculated masses of harmonic quarks 
is estimated as 0.005\%.
\end{abstract}

\section {Introduction}

A many various quark models were applied for investigation of hadrons 
structures~\cite{dalitz}--\cite{collins}.
The part from them used a relativistic approach, i.e. the masses of quarks 
are considered small values or even zero~\cite{ishida}.
A potentials of various forms frequently use, including the harmonic quark 
oscillator and degree functions~\cite{martin, lee, ishida}. 
Others shell quark models were constructed in the no relativistic
limit~\cite{dalitz, isgur, lee, collins}. On this way are received
a good results for the mass spectrum of L=1 baryons~\cite{collins}. 
There are also other non-standard attempts to attract a shell models for 
exposition and systematizations of hadrons~\cite{pallazzi}.

In present work for the first time the rigid system of harmonic quarks and 
model of harmonic potential wells is applied for investigation 
of hadron structures.
In paper~\cite{my1} the simple equation for masses of quarks is obtained on the 
basis of two postulates.

\par {\bf The postulate 1.}
\par {\em A quark-antiquark pair of one flavor~n can annihilate
in part up to a bound state, that defines by the mass equation $m_n\cdot 
{4\over\pi}$.}

We may term this bound state as \textbf {the full harmonic quark oscillator 
or quark-oscillator}.
\par {\bf The postulate 2.}
\par {\em No stable symmetric state harmoniously annihilating oscillator from 
a quark-antiquark pair of one flavor can be broken by that or other way and 
pass at conservation of energy in an unsymmetrical state 
from a quark-antiquark pair of neighboring flavors.}

\par\medskip 
Simplifying the postulate 2, we actually state the following: the total  
mass of two quarks 
with flavors n and n--1 is equal to mass of a harmonic oscillator from 
quark-antiquark pair with flavor n.
From here the simple recurrent equation for quark masses follows:

\begin{equation}\label{MQ}
   {m_n\over m_{n-1}} = {\pi\over 4 - \pi} = 3.659792...\cong 3.66
\end {equation}

The equation can be noted and so:
\begin{equation}\label{qmass}
  m_n = m_0\left[{\pi\over4-\pi} \right]^n
\end {equation} 
where $m_0$ -- mass of a hypothetical quark with zero flavor; $m_n$ -- mass of a 
quark with flavor $n$. \\ 
These equations relate univalently a masses of all quarks in one spectrum and, 
thus, all quarks of 
the harmonic model form the rigid sequence or one set with single "successor" 
in each generation. 
Quarks of the harmonic model have a logarithmic equidistant mass spectrum, and 
their
 flavors are serial numbers of levels in this spectrum or a values n in the 
Eq.~(\ref{qmass}). 
In mathematical aspect there would be enough to postulate an equation for 
quark masses, i.e. Eq.~(\ref{MQ}).

Further the masses of harmonic quarks have been calculated with precision \ApproxP 
{0.03}~\cite{my1}. In table 1 these 
masses of quarks together with energies of their harmonic oscillators 
are given.

\begin{center}
Table 1. The masses of harmonic quarks and their oscillators.  
 
  \medskip
  \begin{tabular}{|l|c|c|c|c|c|c|c|}
    \hline
   n (flavor) & 1 & 2 & 3 & 4 & 5 & 6 & 7\\
    \hline
    Quark & $d$ & $u$ & $s$ & $c$ & $b$ & $t$ & $b'$\\
    \hline
  Quark   & 28.815 & 105.456 & 385.95 & 1412.5 & 5169 & 18919 & 69239\\
mass, MeV &  &  &  &  &  &  & \\

\hline
 Oscillator & 36.69 & 134.27 & 491.4 & 1798.4 & 6581 & 24088 & 88158\\
 energy, MeV & &  &  &  &  &  & \\ 
 
    \hline
  \end{tabular}
\end{center}

Then in~\cite{my1, my2} we have shown on examples 
how the harmonic quarks may be used for description of structures 
and a masses of some particles. So, the strong results of the harmonic 
model is the resolution of a riddle of muon mass and interpretation 
of quark structure of a neutral pion. 
{\textbf {The muon is successful attempt of the Nature,  
suppressing fractional electrical and color charges, 
explicitly to fix one \quark{u}-quark mass 
state as a lepton.
The neutral pion with a large probability is the stationary harmonic 
oscillator from \Quark{u}-antiquark pair}}.
Also there was shown as the harmonic quarks may be used for interpretation
of quark structure of some light mesons and baryon resonances~\cite {my2}.
Since this article, we transfer to systematic investigations of structure of 
hadrons with the help of the harmonic quarks and their full oscillators.

\section {The harmonic quarks in hadrons}

Already in~\cite{my2} it was scored, 
that the structure of a fundamental particle \EM is probably similar to 
structure of atom of helium. It perhaps consists from two filled shells: 
the stationary $s$-quark oscillator and the external light $d\bar{d}$-quark pair. 
The analogy was marked also with helium nucleus, which contain two 
filled shells from two protons and two neutrons.
These analogies are deeply symptomatic and expected.
Atoms and nucleuses with the filled shells have 
the greatest binding energy and stability. 
What should we expect from ensemble of quarks what naturally hadrons are? 
Quarks of hadron, as ensemble of fermions, will be united in some steady 
configurations on the basis of the strong coupling. Whereas steady 
configurations have the minimum energy it is possible to expect, 
that hadrons first of all will contain also the filled shells.
We also shall search for the filled shells in hadronic structures.

\subsection { Filled shells }

The nuclear ensembles of fermions, 
with all definiteness have shown us, that the greatest binding energy and the greatest stability is observed at nucleuses with the filled shells, 
both on protons, and on neutrons. 
These are so-called twice magic nucleuses -- $^4$He, $^{16}$O, $^{40}$Ca. 
The appropriate configurational numbers for nuclear fermions is 2, 2 + 6... . 
At transition from a Coulomb interaction in atoms to a strong coupling in 
nucleuses the considerable 
role on smaller distances of a spin-orbit interaction was found out 
also. 
Whereas a distance between quarks in hadrons is less than a distance between 
a nucleons in nucleuses the role of a spin-orbit interaction can even increase. 
For the further analysis we shall be limited only 
by two filled configurations $s^2$ and $p^6$ which naturally give numbers 
2, 6 and 2 + 6.
In mathematical aspect the configuration $p^6$ is directly 
bound to rotational symmetry of three-dimensional space and 
$s^2$ is bound with dot central symmetry.
In QCD we deal with color interaction which in mesons exists between 
a pair of quarks 
with color and anticolor, i.e. this is a numeral 2.
Baryons have 3 valence color quarks and, in view of 
anticolors, we have a numeral 6. So we see, completely 
electro neutral and colorless groups in QCD, for example, from quarks and antiquarks 
of one flavor can be 2 and 6. That is in the complete correspondence 
with especially stable nuclear configurations.

So, we shall suppose, that completely neutral configurations with numerals 2 and 
6 of quarks of one flavor can form the filled shells with the minimum energy 
and be part of hadrons.
Further we shall suppose also, that the filled $s^2$ and $p^6$ shells can be composed 
from a quark-antiquark pairs which are taking place or in a states 
of a full harmonic oscillators, or in the "free" states. These shells can be 
designated as $0s^2$, $0p^6$, $1s^2$ and $1p^6$.
The energy contribution to a hadron of one pair of quarks in states $0s^2$ and 
$1s^2$ accordingly will be \ApproxP{127} and \ApproxP{200} from mass of one "free" 
quark (Tab.1). Stable two-quark mesons (pions and kaons) place in potential 
wells of full harmonic \quark{u}- and \quark{s}-oscillators~\cite{my1}. 
It allows us to assume that other hadrons are placed also in potential wells 
of full harmonic oscillators or, 
that is more truly, in their combinations. 
The simple potential wells for light and strange 
mesons are already engaged by pions and kaons.
Now, we shall try to construct some system of mesons and then it will be compared 
with experimental spectrum of meson masses.
But before start it is necessary to make some notes.

\subsection {Notes and Designations}

First note is about \quark{d}-quark. This quark should have some special 
properties as against other quarks.
\quark{d}-quark is the most light and therefore most mobile.
It is last quark in a chain of quark decays and their oscillators. 
It may form at decay of \quark{u}-quark or a harmonic \quark{u}-oscillator, 
i.e. it is excitation downwards~\cite{my1}, but itself \quark{d}-quark 
is not capable to generate a new quark with smaller mass.
\quark{d}-quark is as though an appendage or original prolongation 
\quark{u}-quark. 
In our opinion, in view of activity of parallel chain of quarks~\cite{my2}, it 
also is the reason of isospin properties of 
\quark{d}- and \quark{u}-quarks.
So, the mass of \quark{d'}-quark is approximately defined  
by the equation~\cite{Klapdor}: 

$m_{d'} \cong m_d*cos\theta  +  m_s*sin\theta\cong$ 110 (MeV),

where $\theta$ -- Cabibbo  angle. 

This value, as we see, is practically equal to \Quark{d_{\parallel}} mass 
of parallel chain
(105.4 MeV~\cite{my2}) or \Quark{u} mass of main chain. 
The subsequent quarks have the complete gang of properties because they have 
neighbors onto an energy scale both from above and from below.
Properties of the heaviest quark are unknown and also should have features. 
Thus, \quark{d}-quark is not completely independent quark. 
Really, there is no a hadron with mass about 60 MeV (2$m_d $ = 57.6 MeV) 
or less; the lightest hadron -- a neutral pion -- is a harmonic 
\quark{u}-oscillator~\cite{my1}.

The second note concerns the sizes of quarks. At the given work 
we shall not take into account distinction of quark sizes. Nevertheless, 
from approximately equal impulses of quarks in a hadron follows, that 
more heavy quarks should place closer to centre of hadrons.

  Let's enter labels for some quark groups. 
{\bf The quarks of full harmonic oscillators} we shall designate by the same 
symbols, but {\bf located in circles}, for example, \cu\,or \cs; 
while the number, which located in a circle, will inform us about 
{\bf number of harmoniously bound quarks and antiquarks of one flavor}.
So, \ctwou  and 
\csixs, will specify the filled 0$s^2$ and 0$p^6$ shells for {\bf harmoniously 
bound groups} from two \quark{u}- and six \quark{s}-quarks accordingly.
In essence \csixs  group consist of three harmonic oscillators on each of 
axes of space, but, taking into account, that these three quark-antiquark 
of a pairs define 0$p^6$ state, it is important to underline it.
Let's also note, that records such as 2\cu\,and \ctwou  are energy equivalent. 
Numerals 2 or 6 and symbols without circles we shall designate a groups 
of quarks in $s$ 
and $p$ states in which quarks have a complete mass.

In view of the notes made above on restricted properties \quark{d}-quark and 
at absence of a mesons with masses below pion we conclude that quark configurations
exclusively from 
\quark{d}-quarks are unreal and we shall not view them.

\section {General simulation}

At the first stage the scanning simulation of harmonic quark spectrum 
has been carried out with using mesons up to 1000 MeV. 
The purpose of calculation is to define optimal set of harmonic 
quark masses in relation to a spectrum of actual mesons.

Calculations are carried out under following conditions: the \quark{u}-quark 
mass was varied from 70 up to 270 MeV with a step with 0.2 MeV; 
the quantity of mesons is equal 14 (antiparticles are excluded).
On each step we calculated a standing of potential wells and the sum 
of deviations of masses of mesons from the nearest wells. 
Further a root-mean-square variance was calculated and then there were make the normalization on number 
of potential wells on an 
interval 0--1000 MeV, i.e. we took into account the current density 
of potential wells on an energy scale.
The normalizing factor is accepted equal to 1 for the set of the harmonic 
quark masses given in Tab.1.
States 0$s^2$ and 0$p^6$ for us are energy degenerate, therefore the 
configuration 0$s^2$ + 0$p^6$ consist simply from 4 harmonic oscillators.
In result, the model spectrum of potential wells consist exclusively of a sets of 
the harmonic quark oscillators in quantity from 1 up to 4 for two 
flavors n=2 and n=3 (\Quark{u} and \Quark{s}), namely:

\, \, \, 2\cu, 4\cu, 6\cu, 8\cu, 

2\cs, 2\cs + 2\cu, ... , 2\cs + 8\cu, 

4\cs, 4\cs + 2\cu, ... .

The current mass \Quark{u} was varied approximately in 3.7 times that follows 
from a relation~\ref{MQ}. In this case we capture all 
possible mass variants for the harmonic quarks.
The computer calculations, which have been carried out on the complete actual 
set of mesons up to 1000 MeV, have shown, that the masses of harmonic 
quarks calculated by us~\cite{my1} are the optimal, i.e. settle down in a global 
dispersion minimum. Figure 1 shows the dependence of a variance on the current 
mass \quark{u}-quark.
On the dispersion curve (Fig.1) the global minimum is observed 
at mass \quark{u}-quark 105.6 MeV. The same value was calculated in~\cite{my1}
by other way.
Any other set of harmonic quark masses gives a boosted variance.
\par
\begin {figure} [htb]
\begin {center}
\includegraphics [scale =.9] {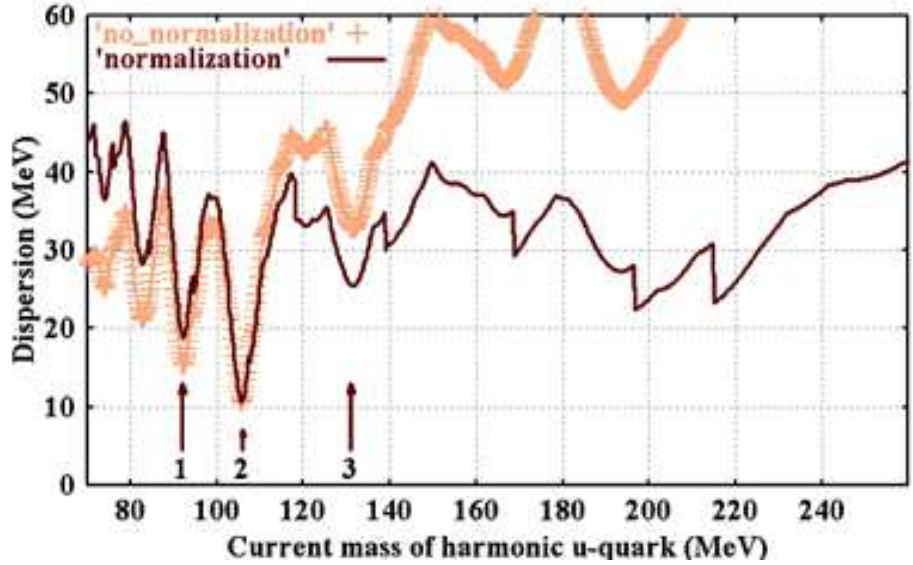}
\par
\caption [] {The change of a variance depending on mass $u$-quark for mesons up to 1000 MeV.
\label {fig:Disp}}
\end {center}
\end {figure}
\par
The dispersion curve has also two local minimums at 92 and 131 MeV.
The study of deviations of meson masses from potential wells for 
minimums has shown, that minimums at 92 and 131 MeV are casual 
events and have no physical content. Figure 2 illustrates concrete values of 
deviations of meson masses from potential wells for according masses of \Quark{u}.

For mass \quark{u}-quark 105.6 MeV a pions and kaons locate into potential 
wells while for 92 and 131 MeV it is not observed.
So, for 92 MeV pions, kaons and $\eta$(548) have a deviation more than 20 MeV. 
For 131 MeV -- pions, \EM, kaons (J=1), \AM and $f$(980).
For \quark{u}-quark with mass 105.6 MeV only $ \omega $ and $\eta$(958) the 
deviation has more than 20 MeV.
\par
\begin {figure} [htb]
\begin {center}
\includegraphics [scale =.9] {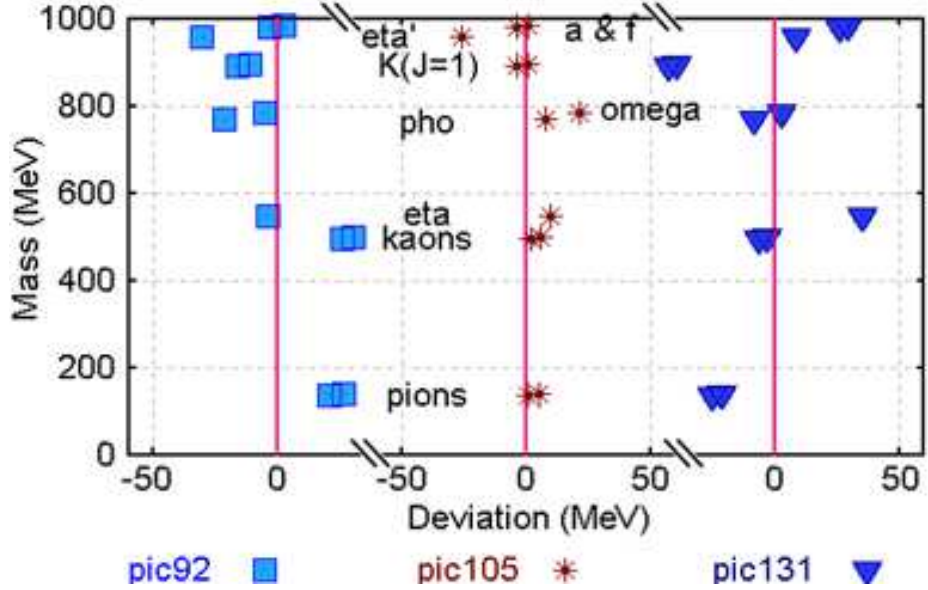}
\par
\caption [] {The deviation of mesons from harmonic potential wells.
\label {fig:Dev}}
\end {center}
\end {figure}
\par
It is important that the global minimum on a dispersion curve keeps 
position at 105 MeV even if to exclude pions from viewing. 
Moreover, its keeps position
even if to exclude both pions and kaons with J=0.
Thus, we have received one more confirmation of validation 
for harmonic quark modell and a uniqueness of mass solutions. 
The dispersion of a global minimum is equal 10.6 MeV.
At "lockout" of an isospin
on average masses of 9 mesons $(\pi, K, \EM, \rho,\,\omega, K$(J=1), 
$\eta (958), \AM, \FM)$
the dispersion of a global minimum is equal 13 MeV and $m_u$ = 105.6 MeV also.

The further simulation was carried out on a generated mass spectrum 
with random distribution of 9 mesons on an interval 100--1000 MeV.
In this case the probability of observation of a variance equal 
or smaller 13 MeV are lower than 4\%. 
Hence, we may say, that the found correlation between the masses of
harmonic quarks~\cite{my1} and an actual spectrum of meson masses  
up to 1000 MeV with probability 96\%, at least, is not casual.
Certainly the actual spectrum of mesons is not casually the distributed and 
at generation of meson masses we should take into account most 
simple requirements, for example, obligatory presence about first 
potential well at least one generated meson or to take into account mass ratios 
for mesons with simple quark compositions, for example, the charged pion 
and kaon.  
Therefore, the probability of casual correspondence of the harmonic quarks and 
actual meson spectrum should be much lower than 4\%. So, only at the fixed 
generation of the second meson (analog $K$) at a potential well 2\cs 
\, with a deviation no more than 5 MeV the probability of casual observation of 
the detected legitimacy decreases with 4\% down to 0.1\%.
If analogs $K$ and $\pi$ is generated within $\pm5$ MeV about wells 2\cs\, and 2\cu , 
then the probability decreases with 4\% down to 0.001\%, 
i.e. we come near to a level of reliability in 4$\sigma$. 
Thus, we can state, that a conformity of real and model mass spectrums are observed
with probability near to 0.9999. 
Figure 3 illustrates this physical legitimacy.
\par
\begin {figure} [htb]
\begin {center}
\includegraphics [scale =.8] {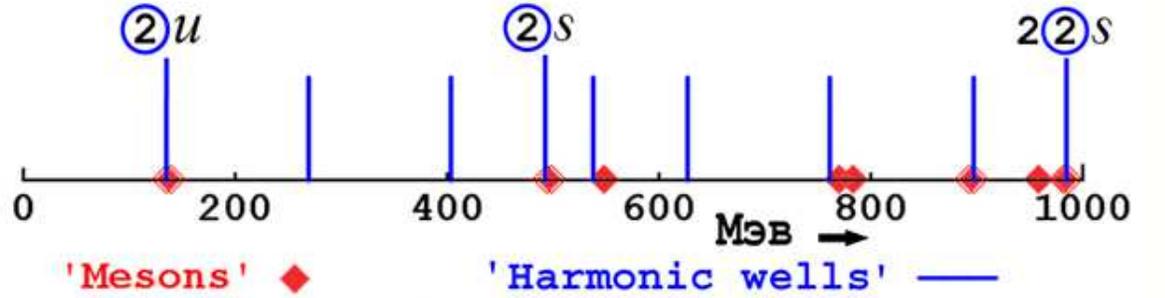}
\par
\caption [] {The arrangement of the model harmonic 
wells at mass \quark{u}-quark 
105.6 MeV and mesons up to 1000 MeV. \label {fig:Sit}}
\end {center}
\end {figure}
\par
Let's estimate degree of correspondence of two spectrums by other way. 
On an interval 1000 MeV the mean distance between wells is \Approx 110 MeV and, 
hence, the mean distance from any position on energy scale to a nearest well 
is lower than \Approx 55 MeV. 
For only random distribution on an interval 0--55 MeV a deviations of 
meson masses from wells would be more or less homogeneously. 
However all 9 deviations is less than half of it interval and the majority 
of them does not exceed 5 MeV. Allocation is completely one-sided. 

The ratio of an actual deviation of meson mass  
to all interval 55 MeV is a probability $P_n$ what a casual event will take place
in a deviation area. Mathematical expectation for $P_n$ is equal 0.5.
On 9 mesons the probability of casual correspondence $P_{c}$ will equal:
\begin{equation}\label{Pc}
   {P_c} = {\prod_{n=1}^9}{P_n\over2} \cong 5\cdot10^{-7}
\end {equation}

Certainly, negligible quantity $P_c$ speaks to us what both meson and model mass 
spectrums are very strongly connected.
Thus we have received the proof of real existence the harmonic quarks and 
oscillators. Besides it is obviously that the harmonic quark masses 
are calculated enough exactly. 

\section {Possible structures of mesons up to 1000 MeV}

Now we can begin detailed study of possible structures of mesons.
Apparently (see Fig.~3) mesons place at the following potential wells:

pions and kaons at single oscillators -- \ctwou\ è  \ctwos;

\EM at \ctwou + \csixu; \,\RM  at  \ctwos +  2\ctwou; \,$K^*$ at \ctwos + \csixu;
\,\AM è \FM at 2\ctwos.

\subsection { One filled harmonic shell }

So, we shall begin with study of the filled configurations with two and six 
harmoniously bound \quark{u}- and \quark{s}-quarks.
We could expect presence of potential wells not only for \ctwou  and 
\ctwo$_s$, but also for the following simple quark combinations:

\csixu  and \csix$_s$.

The depth of potential well of separate harmonic oscillator \ctwou  and \ctwos 
is accordingly equal 77 and 
280 MeV.

In Table 2 these quark configurations are given together with a mesons, which 
is in these potential wells.
\begin{table}
\begin{center}
Table 2. The energies of simple harmonic quark combinations \\
	and masses of some mesons.
  \medskip\small
 \begin{tabular}{|c|c|l|c|c|}
    \hline
Quark & Configurational & Meson & Mass of a meson, & Difference, \\
configuration & energy, & & MeV & MeV \\
        & MeV & & &  \\
   \hline
    \ctwou & 134.27 & \pion & 134.98 & 0.71 \\
  &  & \meson{\pi}  & 139.57 & 5.30 \\
\hline

    \csixu & 402.81 &  - & - &  - \\
\hline

   \ctwos & 491.4 & \meson{K} & 493.65 & 2.25 \\
    &  & \kaon & 497.67 & 6.27 \\
\hline

\csixs &1474.2&$a_0$(1450)&1474$\pm$19&-0.2 $\pm$ 19 \\
\hline

  \end{tabular}
 
\end{center}
\end{table}
Data in Table 2 show, that in potential wells with one quark oscillator 
are placed a stable mesons of a hadronic spectrum: pions and kaons.
That's right. In potential wells with 
three oscillators we find only one meson triplet $a_0$(1450).
The configuration \csixs  is filled shell of type 0$p^6$. 
We could count that a configuration \csixs is far-fetched. 
It does not give anything essential, 
and detection of $a_0$(1450) in area \csixs is simple 
concurrence. However we should remember, that in a potential well \textcircled 
{\footnotesize 4}\quark{s} with unfilled shell 0$p^4$ and with energy 982.8 
MeV we also detected a triplet 
$a_0$(983) and meson $f$(980).
Nevertheless, in the area of energy \csixu there 
are no mesons, and we shall try to understand, why it happened.

Configurations \csixu and \csixs differ from each other by energy and number of 
accessible flavors. At the energy of a configuration \csixu a mesons 
can consist only from \quark{d}- and \quark{u}-quarks. 
Except for the filled shell $0p^6$, consisting of three 
harmonic \quark{u}-oscillators, one more configuration 
with the same energy and the filled shells 
is possible:
0${s^2 (\cu) + 1s^2 (\quark{u}) + 1s^2 (\quark{d})}$. 
Nothing hinders both configurations
immediately to break up to three 
harmonic oscillators, if energy of \csixu is higher than mass of three $\pi^0$.
Any potential barrier at this decay should not be because of a special standing 
\pion in hierarchy of mesons.
As we already have assumed earlier in~\cite {my1} and mentioned above, \pion is 
easiest stable hadron, probably, the unique hadron, which have structure only 
from one stationary harmonic oscillator. 
All other applicants with quarks of other 
flavor, for example, \kaon and $D^0$ have neutral antiparticles. 
So, \kaon makes an attempt to turn into the state of harmonic 
oscillator. However this state is unstable for \kaon
and with probabilities 50\% \kaon is returned in a former state or 
in a state of an antiparticle. 
In other words, in our opinion, the 
reason of known oscillations \kaon is connected with its unsuccessful attempts to 
form a stationary harmonic oscillator. Neutral \pion is unique in this 
respect.
It simply has no opportunity to turn into any other quark state as against  
\kaon ($s\bar{d}$ or $\bar{s}d$). This unique standing \pion 
probably also is the reason of absence of a neutral meson 
in a potential well with a configuration \csixu as it is separated 
without a barrier on 3 stable 
\quark{u}-oscillator (3\pion).
Besides if a real level 0$s^2 $ is disposed below the levels 0$p^6 $, 
then the filled configuration from two shells will contain 
4 \quark{u}-oscillator. It would mean, that real meson may have the mass 
a little more 537 MeV.

Area of energies \csixs is richer by both an energy opportunities and  
a quark flavors. 
Here depth of a potential well is much greater therefore
existence of a meson is quite probable.
Without taking 
into account \quark{d}-quarks two configurations with the filled shells and with 
energy 1474 MeV are here also possible:

\csixs  \, and  \, \ctwos + $s\bar{s}$ + $u\bar{u}$. \\   
Last configuration will consist of three filled shells of type $s^2$:

the first shell from the most bound state, e.g. two harmonic \quark{s}-quarks;

the second shell from the "free" pair \quark{s}-quarks; 

the third shell from the "free" pair \quark{u}-quarks. \\
At the account \quark{d}-quarks the number
of the filled configurations will considerably increase. 
Decay of this 
configuration on 3 strange of an oscillator does not occur because 
there is no the strange analog $\pi^0$. 
The strong decay on 3\kaon is prohibited and as a 
result in the well we have a resonance $a_0$(1450).

\subsection { Two filled harmonic shells }

Now we shall transfer to the analysis of quark configurations with two filled 
shells. Table 3 demonstrates these combinations. They were composed 
of simple wells which are given in Table 2. 
\begin{table}
\begin{center}
{Table 3. Composite combinations from harmonic oscillators and masses \\
		of some mesons.}

  \medskip\small
 \begin{tabular}{|c|c|l|c|c|}
    \hline
Quark & Configurational & Meson & Meson mass, & Difference \\
configuration & energy, & & MeV & of masses, \\
 & MeV & & & MeV \\
    \hline
    \ctwou\ + \csixu & 537.1 & \EM & 547.7 & 10.6 \\    
    \hline
    \hline
    \ctwou + \ctwos & 625.7 & - & - & - \\
    \hline

    \csixu + \ctwos & 894.2 & $K^{*\pm}$ & 891.7 & -2.5 \\
	& & $K^{*0}$ & 896.2 & 2.0 \\

\hline

   \ctwou + \csixs  & 1608 & $\pi_1$ & 1593$\pm$8$\pm$32 & -15$\pm$8$\pm$32 \\
    &  & $X(1600)$ & 1600$\pm$100 & -8$\pm$100 \\
\hline

    \csixu + \csixs & 1877.0 & $p\bar{p}$ & 1876.54 & -0.46$\pm$0.56 \\
\hline
    \ctwos + \csixs & 1965.6 & $K_2^*$& 1973$\pm$8$\pm$25& 7.4$\pm$8$\pm$25 \\
	& & $K^0$ & 1945$\pm$10$\pm$20 & -19.4$\pm$10$\pm$20 \\
	& & $X(2000)$ & - & - \\
	& & $f_2(1950)$ & - & - \\

\hline

\end{tabular}
 
\end{center}
\end{table}
Even at absence of actual mesons in the area of the most light 
mixed configuration \ctwou + \ctwos  the data of Table 3 
bring to us a presents.
At first, in the potential well from two harmonic filled shells
\csixu + \ctwos appeared the first strange vector doublets
$K^{*\pm}$ and $K^{*0}$, and the centre of a well is located 
precisely between mesons. 
Second, energy of a configuration \csixu + \csixs  
is equal precisely to mass of the proton-antiproton pair.
This relation is true with an error of definition of quark 
mass, that was calculated in~\cite{my1} and equal \ApproxP{0.03}.
This error in recalculation on group \csixu + \csixs is given in Table~3 
($\pm$0.56 MeV).
It is more than energy difference between $p\bar{p}$ and harmonic group, 
as this should be at the correct definition of precision with which 
the masses of harmonic quarks are calculated.
The new energy equality gives unique occasion for reflection 
about the nature of the proton and same time we might hope onto 
increasing of precision of quark mass calculation. 
Now we shall sum up the data of Table~3. 
Except for the configuration \ctwou + \ctwos all others 
harmonic configurations with two filled shells successfully work. 
In double potential wells we are discovered an important mesons and even 
unexpectedly $p\bar{p}$ a threshold state.

Thus, we can summarize the achieved results which obtained in Tables 2, 3 
and also~\cite{my1}.
We managed to locate in harmonic potential wells of full quark oscillations 
the following hadrons up to 1000 MeV: 

pions, kaons, \EM, $K^{*\pm}$, $K^{*0}$, nucleons, \FM and \AM. 
\medskip
Besides we shall take into account, that the mass of the 
first vector meson with latent strangeness \OM only on 11 MeV is 
more than mass of the strange pair of quarks $s\bar{s}$ (772 MeV).
The width of a resonance \OM is 8.4 MeV, i.e. peak begins 
immediately at reaching of a threshold for $s\bar{s}$ pair.
So now for completeness of a search we should find 
a place only for two mesons \RM and ${\eta'}$.
To find quark configurations of these mesons we shall try to use 
already found neighboring potential wells with smaller energy. 
For ${\eta'}$ the nearest filled 
configuration is \csixu + \ctwos with energy 894.2 MeV. The difference (64 MeV) 
is close to mass of $d\bar{d}$ quark pair (58 MeV). 
Therefore, the configuration ${\eta'}$ could 
be following: \ctwos + \csixu + $d\bar{d}$. 
Energy of a configuration 
951.9 MeV on 6 MeV is less than the mass of ${\eta'}$.
Then this configuration will consist of three filled shells 
and it is good that two of 
them -- the heaviest and the most light -- same as a quark shells 
of \EM meson. 
Identical external filled shells of both mesons, perhaps, also 
determine their properties and a title, as ${\eta}$ hadrons. 
The accepting this configuration for the true, we could tell, 
that the solution of secret of existence and mass ${\eta'}$ 
was reached only with the help of the harmonic quarks.
We could go farther and to try to define the mass of ${\eta'}$. 
For this purpose let us
to replace the mass of a harmonic oscillator \ctwos on mass \kaon 
and then the mass of ${\eta'}$ should be equal 958.1 MeV.
For such replacement we have some basis.
As we mentioned already \kaon in solo fulfillment make 
unsuccessful tries 
to turn into a state of a stationary harmonic oscillator. 
Presence of other 
shells can stabilize this process. 
The part of excess energy of a kaon about 6 
MeV (over the strange harmonic oscillator) can be redistributed 
to external shells. 
This loss of energy can stabilize the strange oscillator 
on time life of meson. 
As shown in~\cite {my2}, the 
most light \Quark{d} in \meson {\pi} can have a kinetic energy 
in some MeV and redistribution of energy from a shell \ctwos 
to a shell $d\bar{d}$ is quite explained.

We can see, that a shell \csixu, not being implemented in solo variant, 
can be formed at the presence of a potential well of the strange oscillator.
Most likely, in this case, the heavy internal shell \ctwos with a deep 
potential well stabilizes the shell \csixu which consist 
of more light \quark{u}-quarks.
Then both shells is present in $K^*$ and ${\eta'}$ mesons. These mesons
different from kaons and \EM by shell \csixu only.

Now let's consider \RM meson. 
The filled harmonic well
nearest to \RM is single empty configuration \ctwou + \ctwos in Table 3.
The difference of energies between them \Approx 140 MeV is close 
to masses of pions and \quark{u}-oscillator. On Figure 3 we see, that \RM
situated beside with wells \ctwos + 2\ctwou.
Therefore as a first step we can assume this configuration for \RM. 

Energy of a configuration is 760 MeV.
The actual quark configuration \RM can be following for  $\rho^0$ and $\rho^{\pm}$:

\ctwos + $u\bar{u}$ + $d\bar{d}$ \, and 
\ctwos + \ctwou + $u\bar{d}$ (or $\bar{u}d$).\\
Thus, for $\rho^0$ we have a configuration from three filled shells 
$0s^2(\cs) + 1s^2(\quark{u}) + 1s^2(\quark{d})$. \\
The $\rho^{\pm}$ consist from two neutral filled shells $0s^2(\cs) + 0s^2(\cu)$ 
and two valence quarks, as well as in $\pi^{\pm}$.
The interval of possible masses \RM can reach from 
energy of a well 760 up to 776 MeV. This maximal energy can be received
for variant \kaon + 2$\pi^{\pm}$. 
In the latter case \kaon is a mass analog of a 
harmonic oscillator $\cs$, in the same sense, as well as in ${\eta'}$.
Experimental medial mass $\meson{\RM}$ also is in an interval 
from 763 up to 776 MeV since it is depends from conditions 
of observation~\cite{PDG}.

Let's summarize some results of our researches of hadron structures 
up to 1000 MeV.
Using only the filled shells of type $0s^2$ and $0p^6$ from the harmonic 
quarks and their oscillators, we managed to construct system 
of potential wells 
adequate to experimentally observable particles.
The obtained data are shown on Figure~3 and are generalized in Table~4.

\begin{center}
Table 4. Energies and configurations of potential wells and hadrons.

 \medskip\small
 \begin{tabular}{|c|c|l|c|c|}
    \hline
Configuration & Configurational & Hadron & Conventional & Hadron \\ 
of potential & energy, & & or possible & mass, \\
hole	 & MeV & & configuration of hadron & MeV \\

\hline
\ctwou & 134.27 &\pion & \ctwou & 134.98 \\
  &  & \meson{\pi}  & $u\bar{d}$, $\bar{u}d$ & 139.57 \\
\hline
   \ctwos & 491.4 & \meson{K} & $u\bar{s}$, $\bar{u}s$ & 493.65 \\
 & & \kaon & $s\bar{d}$ $\Leftrightarrow$ \ctwos\, $\Leftrightarrow$ $\bar{s}d$  & 497.67 \\
\hline
 \ctwos & 491.4 & \EM & ?!  \ctwos\ + $d\bar{d}$ & 547.7 \\

\ctwou\, + \csixu & 537.1  &  &  2($u\bar{d}$ + $d\bar{u}$) &  \\
\hline
\ctwos\, + 2\ctwou & 760 & $\rho^0$ & \ctwos\, + $u\bar{u}$ + $d\bar{d}$ & \Approx769 \\
	& & $\rho^\pm$ & \ctwos+\ctwou+($u\bar{d}$ or $\bar{u}d$) & \Approx769 \\
\hline
 - &  2$m_s$ \Approx772  & \OM & $s\bar{s}$ & \Approx783 \\
\hline
\ctwos\ + \csixu  & 894.2 & $K^{*\pm}$ & ($u\bar{s}$ or $\bar{u}s$) + \csixu  & 891.7 \\
	& & $K^{*0}$ & ($s\bar{d}$ or $\bar{s}d$) + \csixu & 896.2 \\
	& & $\eta'$ & \ctwos\ + \csixu\ + $d\bar{d}$ & 958 \\
\hline
\cthreeu\ + \cthrees & 938.5 & $p$ or $\bar{p}$ & ? ? ? & 938.3 \\
\hline
2\ctwos & 982.8 & \AM & \RM + $u\bar{u}$ &\  984.7 \\
	& & \FM & $s\bar{s}$ + $u\bar{u}$ & \Approx980 \\
\hline

\end{tabular}
 
\end{center}
All hadrons up to 1000 MeV are submitted in Table 4.
Two groups of mesons pay attention pays to itself.
These groups are very similar each other under the 
status of particles:
\begin {itemize}
\item \meson {K}, \kaon and $\eta$(548);
\item $K^{*\pm}$, $K^{*0}$ and $\eta'$.
\end {itemize}  

Both groups differ only by the filled neutral shell \csixu which 
does not change accessories of mesons to the strange particles or $\eta$.
Besides on an example of $\eta$-mesons we may note that the potential 
well is capable to keep an additional quark-antiquark pair.
But we should remember, that there are variants of decisions 
for $\eta$ mesons. 

\subsection { Proton-antiproton configuration }

Certainly, the most important and unexpected result is obtained 
for the pair of proton-antiproton. 
Decay of a configuration \csixu + \csixs in halves on  
$p\bar{p}$ pair can be accomplished by various ways
with formation completely vague 
inside baryon structures.

However, it is very important, that decay happens of two 
neutral boson configurations $0p^6$, each of which consists of three pairs 
of full harmonic oscillators.

One possible variant of decay of the symmetric configuration is 
submitted on Figure 4. The decay generates a proton consisting of 
three \textcircled{\footnotesize {\em u}}-quarks and 
three \textcircled{\footnotesize {\em s}}-quarks so that the total 
electrical charge is equal +1.

\par
\begin {figure} [htb]
\begin {center}
\includegraphics [scale =.85] {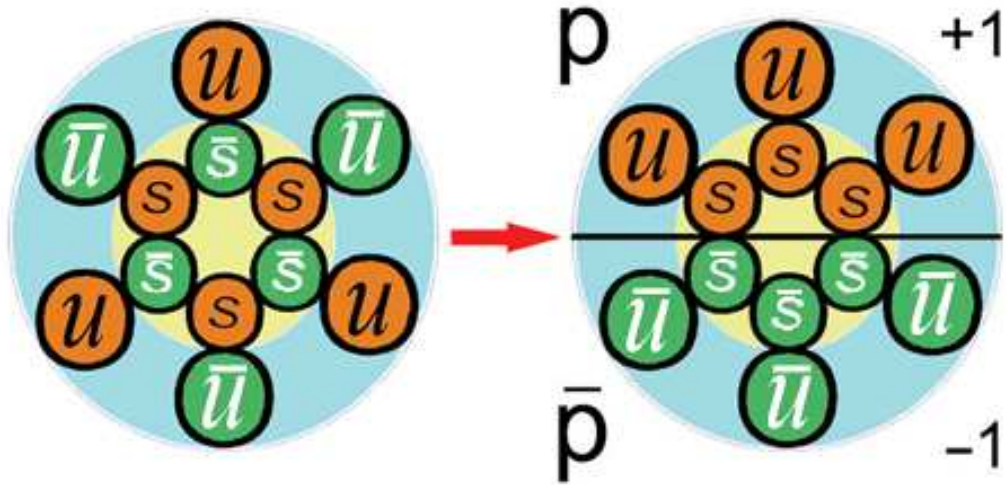}
\par
\caption [] {The decay scheme of two harmonic $0p^6$-configurations 
on a proton and an antiproton.
\label {fig:Proton}}
\end {center}
\end {figure}
\par
If the main rules of QCD for harmonic bound quarks is same as for free quarks,  
i.e. their electrical and color charges is equal, then
 we discover, that the variant of decay 
on Figure 4 is unique. 

If to unite \cu\, and \cs\, in the pairs \textcircled{\scriptsize us}, we can 
receive three partons with a charge +1/3.
Let's mark, that in ÊÕÄ such decay will give in doubling color charges 
in a proton, as against the concept conventional now.
In too time this doubling of color charges gives only the doubling of the total 
white charge, i.e. this consequence is superable.
On the other hand, we can suppose, that one of two groups of a color
charges annihilate at decay and at formation of partons, then the color 
of a proton remains white in the conventional sense.

Because of the complete vagueness of a problem in Table 4 for a proton and an 
antiproton is conditionally written down the following configuration:
\textcircled{\footnotesize 3}$_u$ + \textcircled{\footnotesize 3}$_s$.

\subsection { Other boson configurations }

Discovery $p\bar{p}$ boson configuration consisting of two filled shells 
$0p^6(u)$ and $0p^6(s)$, has allowed us to assume, that perhaps there are others 
important boson configurations, which can generate a hadrons at decay 
at conservation of total energy. 
Three simple groups concern to configurations of kind 1$p^6$:
6\quark{u}, 6\quark{s} and 6\quark{u} + 6\quark{s} with energies 632.7, 
2315 and 2948 MeV.

These electro neutral configurations are also completely colorless groups 
since each shell $1p^6$ have three colors and three anticolors.

The first configuration, we shall term it {\bf 6\quark{u}-boson}, presents to us 
a new very important equality of energies.

It is easily to be convinced, that the sum of masses 
of two most light neutral mesons \pion and \kaon (632.649 $\pm$ 0.03 MeV)
is exactly equal the mass of 6\quark{u}-boson.
\pion is truly neutral particle. 
\kaon in long-term (average integrated) sense is truly neutral 
particle because it is capable to turn into of a state a particle to a state 
an antiparticle and on the contrary.
Thus, we have energy equality of masses of two truly neutral groups. 
Probability of casual concurrence of energies is close to zero 
because below 632.7 MeV there are only two simple neutral mesons \pion and $K^0$, 
which represent two simple harmonic oscillators: 
 \ctwou and \ctwos.
Nevertheless we should estimate a probability of an accidental coincidence 
of these energies. 
Estimation of probability of casual event $P$ it is feasible counting upon 
hit of an integer \quark{u}-quarks on value of the sum of meson masses 632.65 MeV:

\begin{equation}\label{MQ4}
	{P \cong 2{\Delta_{6\quark{u}}\over m_u} \cong 0.003}
\end{equation}
where $\Delta_{6\quark{u}}$ -- the error with which 6$u$-boson mass was defined.
Complementary probability (0.997) says to us about some  
objective order of micro world that gives raises this equality.
 
All this series of important "coincidences" testify only that our 
deductions are true and a manifestation of the harmonic quarks have a 
true systematical character 
($\mu, \pi^0, K^0, \eta(548)$, $K^*$, \DR~[2], $p\bar{p}$, etc.).
If we shall estimate the total probability of all casual coincidences on all this  
array that a result shall completely insignificant and further we will not more 
return to their discussion.
Practically all hadrons up to 1000 MeV concern to L=0 background states 
and the author is sure, that the harmonic quarks and their full oscillators 
are the tool for the analysis and understanding of these states.
 
The author believes in possibility, that a found 
energy equalities are real manifestation of the mechanism 
of a mass-formation in Higgs sector. 

\subsection { Masses of harmonic quarks } 
 
So, we have two important equalities:

\begin{equation}\label{PP}
   m_{p\bar{p}} = 3\cdot{4\over\pi}(m_u + m_s) 		
\end {equation}

\begin{equation}\label{PIKA}
m_{\pion} + m_{\kaon} = 6m_{\quark{u}}	 		
\end {equation}

We yet do not understand sense and meaning of equalities (\ref{PP}) 
and (\ref{PIKA}). However 
with believing, that these equalities are true and reflect certain actual 
physical processes, we can use 
these equalities for improvement of masses of the harmonic quarks. 
As the precision of total mass of  \pion and \kaon exceeds the calculated precision of mass 6\quark{u}-boson 
(632.736$\pm$0.18) in 6 times then this give a chance to improve an exactness of quark mass definition up to $\pm$ 0.005\% (see Table~5).

\begin{center}
  Table 5. The precise masses of harmonic quarks (MeV)
  \nopagebreak\par\medskip
  \begin{tabular}{|l|cccccc|l|}
    \hline
    Boson& $d$ & $u$ & $s$ & $c$ & $b$ & $t$ & error,\%\\
    \hline
    \meson{b}[1]&28.815&105.456&385.95&1412.5&5169.4&18919&$\pm$0.030\\
\hline

    \pion + \kaon&28.811&105.441&385.89&1412.3&5168.7&18916&$\pm$0.005\\
\hline
   
 $p\bar{p}$&28.807&105.429&385.85&1412.1&5168.1&18914& ??? \\
    
    \hline
  \end{tabular}
\end{center}

The distinction of quark masses, which was calculated 
of two new relations (\ref{PP}) and (\ref{PIKA}), 
is only 0.01\%, but nevertheless it exceeds in two time 
a relative error of mass definition  of group \pion + \kaon.
In spite of the fact that the relative precision of 
proton mass definition is better on the one order, than for a kaon, 
nevertheless the difference in \ApproxP{0.01} is referred to a proton.
For this reason, 
in Table 5 in the cell "error" for a $p\bar{p}$ pair are put the symbols ???.
This care has some basis because a proton is a stable unique fermion 
with the considerable magnetic moment as against mesons \pion and $K^0$.
Therefore the decay of configuration \csixu + \csixs  probably 
may be accompanied by some 
allowances to mass of a proton. 
At present we do not know neither type of an allowances, nor their 
quantities.
Existence of the anomalous magnetic moments at an 
electron or a muon is the quite sufficient reason for caution.
Having it in a view, the author recommends using the values of 
quark masses, obtained of the mass sum $m_{\pion} + m_{\kaon}$.

\medskip
From mass relation (\ref{PIKA}) the important deduction follows about a 
possible configuration of \RM meson. 
We built the its configuration on the basis of a potential well 
\ctwou + \ctwos with addition of a pions. 
But the potential well \ctwou + \ctwos in actual meson fulfillment is essence 
\pion + $K^0$. 
Taking into account a relation (\ref{PIKA}) we can note the new variant 
of configuration 
for $\rho$-mesons:  pion  +  6$u$-boson or  pion  + 1$p^6$(u). 
Thus, we have received for $\rho$-meson a configuration of a pion with an 
additional filled shell 
1$p^6$(u), which is completely electro-neutral and colorless.
At present we have no answer to a question: is which implemented 
of these two configurations? 
Nevertheless last configuration seems more preferable with both on  
simplicity (only two shells) and on likeness to a triplet of pions.
Besides this variant has a likeness of shells 
with the vector mesons $K^*$.

In the end let's consider last configuration 6\quark{s} with energy 2315 MeV. 
Directly in this area of energy we have three $f$-mesons: $f_2$(2300), 
$f_4$(2300) and $f_2$(2340).
For a configuration 6\quark{s} the role similar 
\pion in group 6\quark{u} plays a harmonic oscillator \ctwos or an actual 
meson $K^0$, which tightly the bound with this oscillator.
Then energy of the second participant will be \Approx 1817 MeV. In this area
we shall detect again $\pi^0$(1800) with J=0 also and $f_2$(1810).
It is not so clear, but encouraging results.

 We shall not view the configuration 6\quark{u} + 6\quark{s} 
 with energy 2948 MeV, 
since it in essence a combination of two above considered configurations. 
The half of this configuration, i.e. 3\quark {u} + 3\quark {s}, is  
identical energetically to group \csixs and is submitted in Table 2.

\end {document}